# Quantifying Efficiency Loss of Perovskite Solar Cells by a Modified Detailed Balance Model

*Wei E.I. Sha#, Hong Zhang#, Zi Shuai Wang, Hugh L. Zhu, Xingang Ren, Francis Lin, Alex K.-Y. Jen, and Wallace C.H. Choy\**

W. E. I. Sha, H. Zhang, Z. S. Wang, H. L. Zhu, X. G. Ren, Prof. W. C. H. Choy
Department of Electrical and Electronic Engineering, The University of Hong Kong, Pokfulam Road, Hong Kong, P. R. China

Email: weisha@zju.edu.cn (Sha); chchoy@eee.hku.hk (Choy).

F. Lin, Prof. A. K.-Y. Jen
Department of Materials Science & Engineering, University of Washington, Seattle, Washington 98195, United States.

Prof. A. K.-Y. Jen
Department of Physics and Materials Science, City University of Hong Kong, Kowloon Tong, Hong Kong.

# Contributed equally to the work

**Abstract**

A modified detailed balance model is built to understand and quantify efficiency loss of perovskite solar cells. The modified model captures the light-absorption dependent short-circuit current, contact and transport-layer modified carrier transport, as well as recombination and photon-recycling influenced open-circuit voltage. Our theoretical and experimental results show that for experimentally optimized perovskite solar cells with the power conversion efficiency of 19%, optical loss of 25%, non-radiative recombination loss of 35%, and ohmic loss of 35% are the three dominant loss factors for approaching the 31% efficiency limit of perovskite solar cells. We also find that the optical loss will climb up to 40% for a thin-active-layer design. Moreover, a misconfigured transport layer will introduce above 15% of energy loss. Finally, the perovskite-interface induced surface recombination, ohmic loss, and current leakage should be further reduced to upgrade device efficiency and eliminate hysteresis effect. The work contributes to fundamental understanding of device physics of perovskite solar cells. The developed model offers a systematic design and analysis tool to photovoltaic science and technology.

**1. Introduction**

Due to merits of direct band gap [1-4], high and balanced carrier mobility [5, 6], long electron-hole diffusion length [7-9], low non-radiative Auger recombination [10], and high internal quantum efficiency [11], perovskite solar cells show a great potential to be the next-generation high-performance photovoltaics. A great amount of schemes have been proposed to enhance device performances,[12-22] and promote power conversion efficiency (PCE) up to 22% [23]. However, limited works discuss the loss mechanism and quantify the efficiency loss for perovskite solar cells.

The analysis and quantification of efficiency loss of solar cells can be done by the drift-diffusion model [24, 25], circuit model [26, 27], and detailed balance model [28, 29]. Due to a high nonlinearity of coupled equations and a complex device configuration, drift-diffusion model is difficult to retrieve simulation parameters from the measured current density-voltage (J-V) curves. The parameters include recombination rate, mobility, energy levels (e.g. conduction band, valence band and work function), effective density of states, and injection/extraction barrier heights. Furthermore, the drift-diffusion model is nontrivial to describe the photon recycling effect correctly [30]. Concerning the circuit model, it requires to retrieve five parameters in modeling. Hence, the uniqueness of the parameters is questionable. Also, the ideality factor in the circuit model has ambiguous physical meaning and cannot quantitatively represent radiative recombination (photon recycling) and non-radiative (Auger and Shockley-Read-Hall) recombination separately. Additionally, while the traditional detailed balance model demonstrates a strong capability to predict the efficiency limit of an ideal solar cell, the model cannot quantify the energy loss of a practical solar cell including optical and electrical (thermodynamic) loss. In this work, a revised detailed balance model is proposed to unveil the loss mechanism and quantify the loss factors of perovskite solar cells. Through investigating the device performance of various fabricated perovskite solar cells, the three dominant loss factors of optical loss, non-radiative recombination loss, and ohmic loss are identified quantitatively. The perovskite-interface induced surface recombination, ohmic loss, and current leakage are also analyzed. Consequently, the work offers a guideline to the researchers for optimizing perovskite solar cells and ultimately approaching the Shockley-Queisser limit of photovoltaics [29].

## 2. Results and Discussion
### 2.1. Theory

In order to understand the loss mechanism and qualify loss factors, we propose the revised detailed balance model, which is expressed as

$$J = \frac{V - JR_s}{R_{sh}} + J_n(V - JR_s) + J_r(V - JR_s) - J_p \tag{1}$$

where $V$ is the applied voltage of a solar cell. $J_p$ is the photocurrent. $J_r$ and $J_n$ are the current loss due to the radiative emission by photon recycling and the non-radiative recombination by defects and impurities, respectively. $R_s$ is the series resistance, which describes the ohmic loss by the contacts, carrier transport layers, and the hetero-junction interfaces between the perovskite and carrier transport layers. The defects and voids induced current leakage is represented by the shunt resistance $R_{sh}$.

The photocurrent is given by

$$J_p = q \int_0^\infty \alpha(\lambda, L) \frac{\Gamma(\lambda)\lambda}{hc_0} d\lambda \tag{2}$$

where $c_0$ is the speed of light in air, $\Gamma$ is the AM 1.5 G spectrum of Sun, $\lambda$ is the wavelength and $q$ is the elementary charge. The absorptivity $\alpha$ is the ratio of power absorbed by the perovskite active layer over the power of incident Sunlight, which depends on the thickness of the perovskite layer $L$, refractive indices of materials adopted, and light-trapping structures. It can be obtained by numerically solving Maxwell equation [31].

The radiative current is written as

$$J_r(V-JR_s) = J_0^r \left[ \exp\left( \frac{q(V-JR_s)}{k_B T} \right) - 1 \right] \quad (3)$$

where $k_B$ is the Boltzmann constant and $T$ is the Kelvin temperature. Here, the radiative saturation current $J_0^r$ is of the form [28]

$$J_0^r = q \int_0^\infty \alpha(\lambda, L) \frac{\Gamma_0(\lambda)\lambda}{hc_0} d\lambda \quad (4)$$

It is proportional to the spectral overlap integral between the absorptivity $\alpha$ and black-body (thermal) emission spectrum $\Gamma_0$ at room temperature ($T$=300 K).

For perovskite solar cells, the dominant non-radiative recombination is the Shockley-Read-Hall (defect-assisted) recombination [32, 33]. Thus, the bulk non-radiative current reads

$$J_n(V-JR_s) = J_0^n \exp\left( \frac{q(V-JR_s)}{2k_B T} \right), \quad J_0^n = q\gamma n_i L \quad (5)$$

where $\gamma$ is the non-radiative recombination rate depending on the crystalline quality of the perovskite film, and $n_i$ is the intrinsic carrier density of bulk perovskite material. The equation (5) also can be used to describe the surface non-radiative current if one replaces $n_i L$ by an equivalent surface intrinsic carrier density.

The modified detailed balance model is different from existing solar cell models (See **Supplementary Note 1**) in literature. Firstly, it is much simpler than the drift-diffusion model and considers the photon recycling effect. Secondly, compared to the circuit model, in the revised detailed balance model, the radiative saturation current is not an unknown parameter needed to be retrieved from experimental data. The radiative saturation current can be calculated with Maxwell's equations and black-body radiation law. Meanwhile, the non-radiative recombination is fully considered in the modified model. Contrarily, it is implicit and inaccurate in the circuit model, which is expressed as the ideality factor.

With the proposed model, (1) we can capture optical effects including light trapping and angular restriction in terms of the absorptivity α and thermal emission spectrum $\Gamma_0$ of perovskite cells. (2) We can describe complete recombination mechanisms including radiative and non-radiative recombination. (3) We can investigate the influences of carrier transport layers and contacts on the electrical response of perovskite solar cells. (4) We can quantify the loss factors in perovskite solar cells involving optical loss, recombination loss, ohmic loss (by series resistance), and current leakage loss (by shunt resistance).

Given a *J-V* curve, our model will adopt the nonlinear Newton's method [34] to retrieve only the three parameters including the non-radiative recombination rate *γ*, the series resistance $R_s$, and the shunt resistance $R_{sh}$. The intrinsic carrier density of $CH_3NH_3PbI_3$ ($MAPbI_3$) material is $10^9/cm^3$ [5]. The refractive index of the $MAPbI_3$ material is obtained by Forouhi-Bloomer model, which is in well agreement with experimental results measured via other methods such as ellipsometry [35].

## 2.2. Analysis of Experiment Results

**Figure 1** shows the investigated device structure and corresponding energy diagram. The fabrication methods are described in **Experimental Section**. $NiO_x$ and PEDOT:PSS are the hole transport layers (HTL). $C_{60}$ is the electron transport layer (ETL). Bis-$C_{60}$ and zirconium acetylacetonate (ZrAcAc) are employed as the electron-selective interfacial layers at the organic/cathode interface [14]. ITO and Ag layers are the anode and cathode, respectively. The $MAPbI_3$ perovskite films are fabricated by a modified anti-solvent method [14].

### 2.2.1 Active layer thickness

The device structures of $MAPbI_3$ perovskite solar cells with different active layer thickness are given as follows:

ITO/ NiO$_x$ (20 nm)/ MAPbI$_3$ (80 nm)/ C$_{60}$ (50 nm)/ Bis-C$_{60}$ (10 nm)/ Ag (120 nm)

ITO/ NiO$_x$ (20 nm)/ MAPbI$_3$ (200 nm)/ C$_{60}$ (50 nm)/ Bis-C$_{60}$ (10 nm)/ Ag (120 nm)

ITO/ NiO$_x$ (20 nm)/ MAPbI$_3$ (240 nm)/ C$_{60}$ (50 nm)/ Bis-C$_{60}$ (10 nm)/ Ag (120 nm)

**Figure 2** depicts the theoretically fitted and experimentally measured *J-V* curves (with the reverse scanning). The relative fitting error is around 2% for all the cases, which suggests the modified detailed balance model well captures the device physics of perovskite solar cells. However, the traditional detailed balance model that does not incorporate the series and shunt resistances produces noticeable errors as illustrated in **Figure 2(b)**.

**Table 1** lists the retrieved characteristic parameters from the measured *J-V* curves. **Figure 3** shows the quantitative efficiency loss for the perovskite cells with reference to the Shockley-Queisser limit, i.e. short-circuit current (*J*sc) of 25.88 mA/cm$^2$, open-circuit voltage (*V*oc) of 1.31 V, fill factor (FF) of 0.91, and PCE of 30.75%. For achieving the Shockley-Queisser limit, the non-radiative recombination should be ignorable ($\gamma$=0). Meanwhile, the series resistance is zero ($R_s$=0) and the shunt resistance goes to infinity ($R_{sh}\rightarrow\infty$). After carefully studying **Table 1** and **Figure 3**, we can understand the evolution of the loss factors as the thickness of active layer increases.

(1) Since thin perovskite active layer (e.g., 80 nm) has a weak optical absorption, optical loss is a dominant loss factor and thus a low *J*sc value is obtained. The low *J*sc reduces the *V*oc, FF, and PCE [28, 30]. In **Table 1**, the thin perovskite layer exhibits the largest non-radiative recombination rate $\gamma$ owing to the low-quality film formation (with low crystallinity and many grain boundaries)[36] resulting in a significant hysteresis effect (**See Figure S1**). However, the thin-active-layer configuration achieves the smallest non-radiative recombination current (loss) that is proportional to the product of the recombination rate $\gamma$ and the active layer thickness *L*, as presented in Eq. (5). As the thickness *L* increases, the non-radiative recombination loss increases, although the corresponding recombination rate decreases and no hysteresis effect can be observed (**See Figure S1**). Regarding the thicker perovskite layer, the optical loss is still large (around 40%) for the device configuration achieving the maximum PCE. Hence, both light-trapping design and formation of high-quality thicker perovskite film are critically important to boost device performance of perovskite photovoltaics.

(2) Because of well-engineered designs of both electron and hole transport layers, the loss by shunt current can be ignored (with sufficiently large shunt resistances) and the ohmic loss (with small series resistances) is as below as 2%. On one hand, thin perovskite film yields a low photocurrent due to the optical loss. On the other hand, it is clear that thick perovskite film absorbs more light and thus produces a larger photocurrent, while too-thick thickness causes loss of photovoltage, most likely due to the increased non-radiative recombination loss.

(3) The non-radiative recombination loss is boosted when the thickness increases (See **Figure 3**). The increased non-radiative recombination is responsible for the reduction of $V_{oc}$ and FF. Most importantly, the non-radiative recombination is the leading loss mechanism for the cell with the maximum PCE. Therefore, improving the crystallinity[12] and reducing the grain boundaries of thick perovskite films are critically important to fabricate highly efficient perovskite solar cells. Meanwhile, passivation of perovskite grain boundaries has also been demonstrated to effectively reduce the non-radiative recombination [37].

*2.2.2 Carrier transport layer*

The MAPbI$_3$ solar cells with different carrier transport layers are given as follows

ITO/ NiO$_x$ (20 nm)/ MAPbI$_3$ (200 nm)/ C$_{60}$ (50 nm)/ Bis-C$_{60}$ (10 nm)/ Ag (120 nm)

ITO/ PEDOT:PSS (40 nm)/ MAPbI$_3$ (200 nm)/ C$_{60}$ (50 nm)/ Bis-C$_{60}$ (10 nm)/ Ag (120 nm)

ITO/ NiO$_x$ (20nm)/ MAPbI$_3$ (200 nm)/ C$_{60}$ (50 nm)/ ZrAcAc (10 nm)/ Ag (120 nm)

In contrast to the optimized device structure where NiO$_x$, C$_{60}$, and Bis-C$_{60}$ function as HTL, ETL, and electron selective interface layer respectively, one case is to replace NiO$_x$ by PEDOT:PSS and the other is to replace Bis-C$_{60}$ by ZrAcAc. Regarding the ZrAcAc case, the ZrAcAc layer is employed to modulate the interfacial properties of the cathode. Although **Figure 4** shows the theoretically fitted and experimentally measured *J-V* curves (with reverse scanning). **Table 2** lists the retrieved characteristic parameters from the measured *J-V* curves. Interestingly, the ZrAcAc case shows the maximum fitting error of 3.27%, which can be observed clearly in **Figure 4(b)** as compared to **Figures 2 and 4(a)**. The injection barrier with an abnormally large *V*oc of 0.94 V (see Table 2) induced by the ZrAcAc layer makes the modified detailed balance model slightly inaccurate.

The quantitative energy loss for the perovskite cells with different carrier transport layers are illustrated in **Figure 5**. A drastically amplified ohmic loss and shunt current loss are clearly observed. The ohmic loss is around 6%. The current leakage loss even reaches 15% for the ZrAcAc case. The enlarged loss can be understood by the increased series resistance and pronouncedly decreased shunt resistance as listed in **Table 2**. Therefore, an improper design of carrier transport layer will introduce significant loss channels by additional resistances and significantly lower the FF. Moreover, the *J*sc reduction of the PEDOT:PSS case is caused by a larger optical reflectance. It is because that PEDOT:PSS has a larger refractive index than NiO$_x$. The reduced *J*sc will also diminish *V*oc. Additionally, no matter which transport layer is adopted, the optical loss fraction is almost unchanged, which is as high as 40% for the thin perovskite film configuration.

2.2.3 *Overall device performances*

Regarding hysteresis effect, we can compare the retrieved parameters for different scanning scenarios (See Table 3). The relative fitting errors are 3.34% and 1.57%, respectively, for the forward and reverse scanning cases (See **Figure S2 and Figure 4(a)**).

The larger fitting error in the forward scanning indicates that there is an injection barrier, resulting in a larger Voc as listed in Table 2. Simultaneous modifications of all the retrieved parameters ($\gamma$, $R_{sh}$, and $R_s$), together with the injection barrier, unveil a complicated mechanism of hysteresis at the interface between the MAPbI$_3$ and PEDOT:PSS layers. It is well known that ions in perovskite will be driven by the internal electrostatic field of the cell. The ions will migrate in the perovskite layer and accumulate at the interfaces between the perovskite and carrier transport layer (PEDOT:PSS) that blocks ions. The migration and accumulation of ions will in turn induce an inhomogeneous spatial distribution of the internal field, which significantly influences the carrier transport by introducing the injection barrier, modifying the surface recombination ($\gamma$), and induce the surface ohmic loss and current leakage ($R_s$ and $R_{sh}$). The simultaneous evolution of all the characteristic parameters under different scanning directions is caused by a complex nonlinear coupling and a giant mobility (relaxation time) difference between carrier transport and ion migration.

The modified detailed balance model is finally employed to quantify the energy loss of optimized and high-performance perovskite solar cells. **Figure S3** shows the corresponding experimental and theoretical results of *J-V* characteristics, with a very small relative fitting error of 0.48%. Such a small error strongly indicates that the modified model near-perfectly captures the device physics of high-performance perovskite cells free from unwanted injection and extraction barriers. The total thermodynamic loss caused by series resistance, shunt resistance and non-radiative recombination is 75%, in comparison with the optical loss of 25%. The reductions of *J*sc, *V*oc, FF, and PCE with reference to the corresponding Shockley-Queisser limit are 12%, 16%, 15%, and 37%. In contrast to the *J*sc loss, both the Voc and FF loss are mainly from the perovskite interfaces. Thus, suppressing the interface loss by reducing series residence and surface recombination is essential to upgrade the efficiency to approach the detailed balance limit.

## 3. Conclusion

We proposed a modified detailed balance model, which shows significant advantages to quantify and understand the efficiency loss of perovskite solar cells. Our experimental and theoretical results show that for thin-active-layer design, light trapping is needed to reduce the dominant optical loss. For optimum device structures, the well-engineered carrier transport layers and high-quality perovskite film are essential to reduce the dominant ohmic loss and defect/impurity induced recombination loss. Our work offers a simple and reliable modeling methodology and insightful understanding to efficiency loss of perovskite solar cells. The model can serve as a general design tool to analyze other emerging solar cells.

## 4. Experimental Section

*Materials*: All the chemicals and materials were purchased and used as received unless otherwise noted. PEDOT:PSS (Baytron Al 4083) was purchased from H. C. Starck GmbH, Germany. $CH_3NH_3I$ was purchased from Dyesol and used as received. $PbI_2$ (99%) was purchased from Sigma-Aldrich. Bis-$C_{60}$ surfactant is provided by Prof. Alex K.-Y. Jen's group. $NiO_x$ NCs inks were prepared according to our previous reports.[14, 38]

*Device fabrication*: ITO-coated glass substrates were cleaned and then ultraviolet-ozone treated for 20 min. PEDOT:PSS (Baytron Al 4083) was spin-coated with thickness of 45 nm and then dried at 150 °C for 10 min. The $NiO_x$ NCs ink (20mg/mL in DI-water) was spin-coated to obtain 20 nm $NiO_x$ film. The resultant $NiO_x$ films will be used to fabricate devices without annealing process or other treatments. The 1M $CH_3NH_3PbI_3$ solution were prepared by reacting the $CH_3NH_3I$ powder and $PbI_2$ in γ-butyrolactone: dimethyl sulfoxide = 7: 3 (v/v) at 60 °C for 1 h. The perovskite precursor solution was deposited onto a $NiO_x$/ITO or PEDOT:PSS/ITO substrate by a consecutive two-step spin-coating process at 1,000 rpm and at 4,000 rpm for 20 and 40 s, respectively, and 180 μl of toluene was rapidly poured on top of the substrates during spin coating in the second spin stage which is slightly modified method from the reported protocol[14]. The obtained films were dried on a hot plate at 100 °C for 10 min. Perovskite layers with different thickness were obtained by varying the precursor concentration. Subsequently, the $C_{60}$ (20 mg/mL in dichlorobenzene) and Bis-$C_{60}$ surfactant (2 mg/mL in isopropyl alcohol) or Zirconium(IV) acetylacetonate (ZrAcAc) (2 mg/mL in anhydrous ethanol) were then sequentially deposited by spin coating at 1,000 rpm for 60 s and 3,000 rpm for 40 s, respectively.

For the best-performance devices, the perovskite precursor was modified by Cl-doping approach. The $CH_3NH_3I$ (190 mg) was mixed with $PbI_2$ (500 mg) and $PbCl_2$ (30 mg) in anhydrous *N,N*-dimethylformamide (1 ml) by shaking at room temperature for 20 min to produce a clear $CH_3NH_3PbI_3(Cl)$ solution. To deposit perovskite film, the $CH_3NH_3PbI_3(Cl)$ solution was first dropped onto a $NiO_x$/ITO substrate. The substrate was then spun at 5000 rpm and after six seconds anhydrous chlorobenzene (180 μl) was quickly dropped onto the center of the substrate, and dried on a hot plate at 100 °C for 10 min. Subsequently, the PCBM:$C_{60}$ (8+12 mg/mL in dichlorobenzene) and Bis-$C_{60}$ surfactant (2 mg/mL in isopropyl alcohol) were then sequentially deposited by spin coating at 1,000 rpm for 60 s and 3,000 rpm for 40 s, respectively.

Finally, the device was completed with the evaporation of Ag contact electrodes (120 nm) in a high vacuum through a shadow mask. The active area of this electrode was fixed at 6 mm$^2$. All devices were fabricated in glove box.

*Measurement and Characterization:* Solar-simulated AM 1.5 sunlight was generated using a Newport AM 1.5G irradiation (100 mW/cm$^2$), calibrated with an ISO 17025-certified KG3-filtered silicon reference cell. The spectral mismatch factor was calculated to be less than 1%. The *J-V* curves were recorded using a Keithley 2635 apparatus. The refractive indices (n, k) of perovskite were performed under a dark ambient environment by using spectroscopic ellipsometry (Woollam).

**Acknowledgement**

W.E.I.S. and H.Z. contributed equally to this work. This work was supported by the University Grant Council of the University of Hong Kong (Grant 201611159194 and 201511159225), the General Research Fund (Grant 17211916 and 17204117), the Collaborative Research Fund (Grants C7045-14E) from the Research Grants Council of Hong Kong Special Administrative Region, China, ECF Project 33/2015 from Environment and Conservation Fund, and Grant CAS14601 from CAS-Croucher Funding Scheme for Joint Laboratories.

**Figures and Tables**

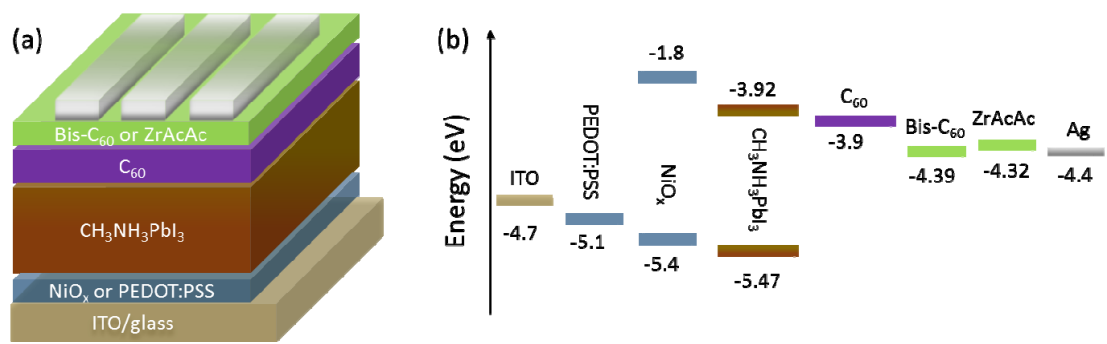

**Figure 1**. (a) Device structure of perovskite solar cells and (b) corresponding energy band diagram.

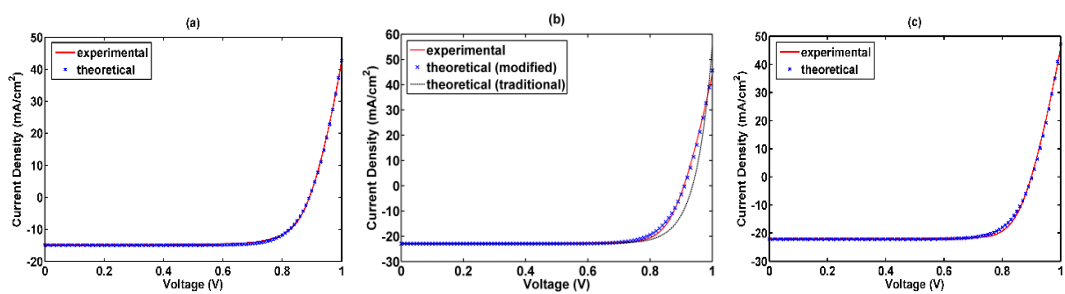

**Figure 2**. The theoretical and experimental *J-V* characteristics of perovskite solar cells with different active layer thickness. (a) 80 nm; (b) 200 nm; (c) 240 nm. The relative fitting errors for the three cases are respectively (a) 1.44%; (b) 2.41%; (c) 2.40%. All the *J-V* characteristics are fitted by the modified detailed balance model. For a comparative study, (b) also presents the result fitted by the traditional detailed balance model.

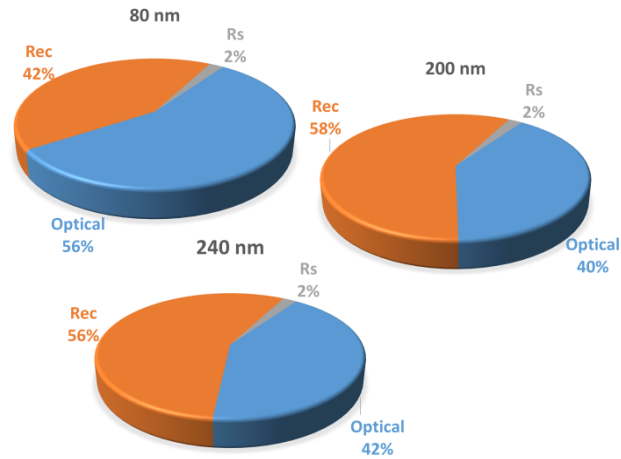

**Figure 3**. Efficiency loss of $CH_3NH_3PbI_3$ solar cells with the active layer thickness of 80 nm, 200 nm, and 240 nm.

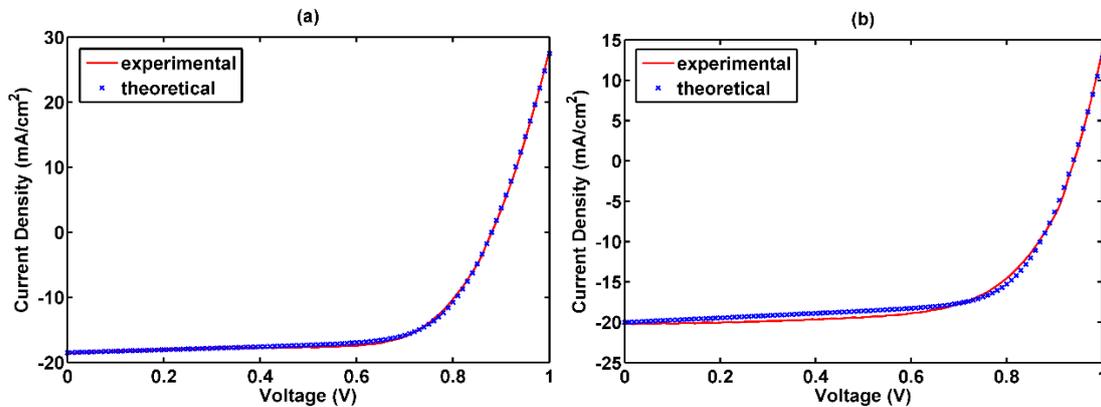

**Figure 4**. The theoretical and experimental *J-V* characteristics for perovskite solar cells incorporating different carrier transport layers. (a) PEDOT:PSS and Bis-$C_{60}$ are the hole transport layer and electron selective interfacial layer, respectively; (b) $NiO_x$ and ZrAcAc are the hole transport layer and electron selective interfacial layer, respectively. The relative fitting errors for the two cases are respectively (a) 1.57%; (b) 3.27%. All the *J-V* characteristics are fitted by the modified detailed balance model.

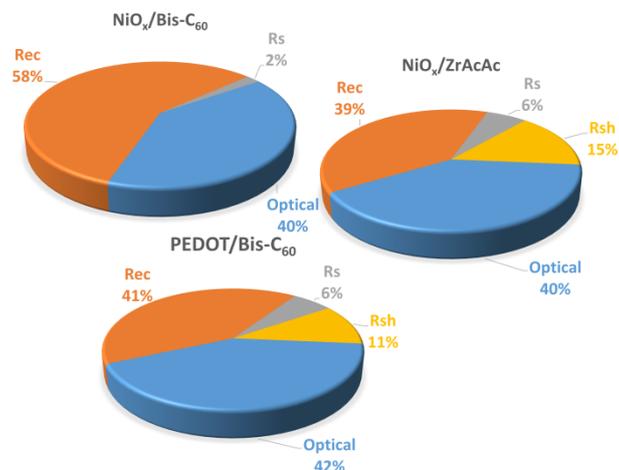

**Figure 5**. The dependence of efficiency loss of $CH_3NH_3PbI_3$ solar cells on the carrier transport layer configurations.

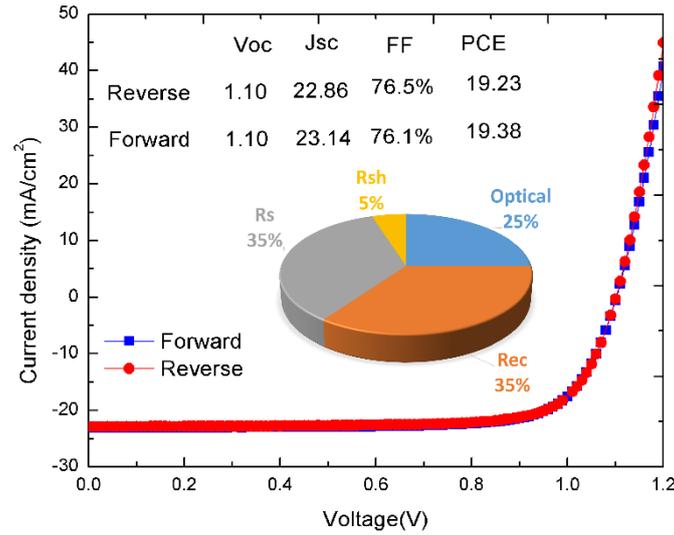

**Figure 6**. Efficiency loss of high-performance perovskite solar cells. The device structure is given as: ITO/ NiO$_x$ (20 nm)/ CH$_3$NH$_3$PbI$_3$(Cl) (300 nm)/ PCBM:C$_{60}$ mixture (50 nm)/ Bis-C$_{60}$ (10 nm)/ Ag (120 nm).

**Table 1**. The retrieved characteristic parameters from measured *J-V* curves of perovskite solar cells with different active layer thickness. *L* means thickness. γ, *R*$_s$ and *R*$_{sh}$ are the non-radiative recombination rate, series resistance, and shunt resistance. The loss of *J*sc, *V*oc, and FF with respect to their Shockley-Queisser limiting values are also calculated.

| *L* (nm) | γ(s$^{-1}$) | *R*$_s$ (10$^{-4}$ Ohm m$^2$) | *R*$_{sh}$ (10$^4$ Ohm m$^2$) | *J*sc (mA/cm$^2$) | *V*oc (V) | FF | PCE (%) |
|---|---|---|---|---|---|---|---|
| 80 | 543 | 0.91 | 0.77 | 14.86/ 43% | 0.89/ 32% | 0.75/ 18% | 9.92 |
| 200 | 228 | 0.71 | 0.85 | 22.97/ 11% | 0.91/ 31% | 0.80/ 12% | 16.72 |
| 240 | 219 | 0.83 | 0.57 | 22.11/ 14% | 0.90/ 31% | 0.80/ 12% | 15.92 |

**Table 2**. The retrieved characteristic parameters from measured *J-V* curves of perovskite solar cells incorporating different carrier transport layers. The loss of *J*sc, *V*oc, and FF with respect to their Shockley-Queisser limiting values are also calculated.

| HTL/interfacial layer | γ(s$^{-1}$) | *R*$_s$ (10$^{-4}$ Ohm m$^2$) | *R*$_{sh}$ (10$^4$ Ohm m$^2$) | *J*sc (mA/cm$^2$) | *V*oc (V) | FF | PCE (%) |
|---|---|---|---|---|---|---|---|
| NiO$_x$/Bis-C$_{60}$ | 228 | 0.71 | 0.85 | 22.97/ 11% | 0.91/ 31% | 0.80/ 12% | 16.72 |
| PEDOT/Bis-C$_{60}$ | 299 | 2.53 | 4.32x10$^{-6}$ | 18.63/ 28% | 0.88/ 33% | 0.69/ 24% | 11.31 |
| NiO$_x$/ZrAcAc | 101 | 2.56 | 3.55x10$^{-6}$ | 20.14/ 22% | 0.94/ 28% | 0.66/ 27% | 12.49 |

**Table 3**. The retrieved characteristic parameters from measured *J-V* curves of perovskite solar cells with PEDOT:PSS and Bis-C$_{60}$ as hole transport and interfacial layers, respectively. The *J-V* curves are measured with the reverse and forward scanning.

| PEDOT:PSS/Bis-C$_{60}$ | γ(s$^{-1}$) | *R*$_s$ (10$^{-4}$ Ohm m$^2$) | *R*$_{sh}$ (10$^4$ Ohm m$^2$) | *J*sc (mA/cm$^2$) | *V*oc (V) | FF | PCE (%) |
|---|---|---|---|---|---|---|---|
| Forward | 115 | 4.32 | 3.47x10$^{-6}$ | 18.65 | 0.93 | 0.63 | 10.99 |
| Reverse | 299 | 2.53 | 4.32x10$^{-6}$ | 18.63 | 0.88 | 0.69 | 11.31 |

# **Supporting Information**

## 1. Supplementary Note 1−Device models for perovskite solar cells

The drift-diffusion model, circuit model and detailed balance model are three commonly-used theoretical models for investigating device physics of perovskite solar cells (PVSCs). Here, we briefly outline the three models.

### a. Drift-diffusion model

The drift-diffusion model is governed by the Poisson, drift-diffusion and continuity equations:

$$\nabla \cdot (\varepsilon \nabla \psi) = -q(p-n)$$
$$\frac{\partial n}{\partial t} = \frac{1}{q}\nabla \cdot \mathbf{J}_n + G - R$$
$$\frac{\partial p}{\partial t} = -\frac{1}{q}\nabla \cdot \mathbf{J}_p + G - R$$

(S1)

where $\mathbf{J}_n = -q\mu_n n \nabla \psi + qD_n \nabla n$ and $\mathbf{J}_p = -q\mu_p p \nabla \psi - qD_p \nabla p$ are the electron and hole current densities, respectively. The electron (hole) diffusion coefficient satisfies the Einstein relation $D_{n(p)} = \mu_{n(p)} k_B T/q$ and $\mu_{n(p)}$ is electron (hole) mobility. Furthermore, $G$ is the generation rate, $R$ is the recombination.

The drift-diffusion model can be used to study the carrier dynamics (transport, recombination and collection) in PVSCs and has the capability to gain the insight of device optimization. However, the predicted results are critically dependent on many empirically/experimentally determined parameters such as mobility, effective density of states, dielectric constant, band gap, work function, recombination rate, etc. In addition, the photon recycling effect cannot be trivially incorporated in this model. Due to a large variation of carrier density (on the magnitude of several orders) and high nonlinearity, a great number of computational techniques have been proposed to solve the drift-diffusion model. However, the retrieval of the simulation parameters from the drift-diffusion model is very difficult.

### b. Circuit model

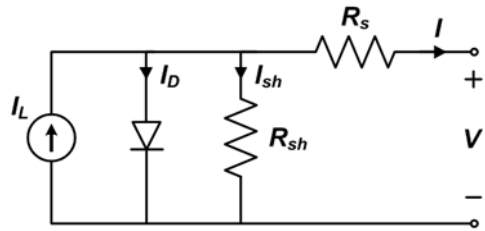

The PVSCs can be modeled by an equivalent circuit above. The current-voltage curve can be expressed by the Kirchhoff's current law for the output current $I$.

$$I = I_{ph} - I_D - I_{sh}$$

(S2)

Here, $I_{ph}$ represents the photogenerated current in PVSCs and $I_{sh}$ is the current flowing in the shunt resistance. $I_D$ is the diode current that can be modeled by the Shockley's equation

$$I_D = I_0 \left[ \exp\left(\frac{q(V+IR_s)}{nk_B T}\right) - 1 \right]$$

(S3)

where $R_s$ and $R_{sh}$ are the series and shunt resistances of PVSCs, respectively. $n$ is the diode ideality factor. $I_0$ is the saturation current, $k_B T$ is the thermal energy at the temperature $T$. Since the $I_{sh}=(V+IR_s)/R_{sh}$, the output current $I$ of the PVSCs under the applied voltage $V$ is:

$$I = I_{ph} - I_0 \left[ \exp\left(\frac{q(V+IR_s)}{nk_B T}\right) - 1 \right] - \frac{V+IR_s}{R_{sh}} \quad (S4)$$

The circuit model provides a simple and fast approach to study the device physics of PVSCs. There are five parameters in the circuit model that needs to be retrieved from the experimental current density-voltage curve, including the saturation current $I_0$, ideality factor $n$, photogenerated current $I_{ph}$, series resistance $R_s$, and shunt resistance $R_{sh}$. As a result, the uniqueness of the retrieved parameters cannot be guaranteed. Moreover, the recombination mechanisms (e.g. radiative, trap-assisted, and Auger recombination) are implicitly involved in the ideality factor $n$. In sum, the circuit model can only give preliminary understandings of device physics in PVSCs.

c. **Detailed balance model**

The photocurrent is calculated as the difference between the photons absorbed by and the photons leaving from the PVSCs:

$$J = J_e(V) - J_{ph} \quad (S5)$$

where $V$ is the applied voltage. $J_e$ represents the current density corresponding to the radiative emission and $J_{ph}$ is the photogenerated current due to the absorption of incident light in perovskite material

$$J_{ph} = q \int_0^\infty \alpha(\lambda, L) \frac{\Gamma(\lambda)\lambda}{hc_0} d\lambda \quad (S6)$$

where $c_0$ is the speed of light in air. $\Gamma$ is the global AM 1.5G spectrum of Sun, $\lambda$ is the wavelength, and $q$ is the elementary charge. $\alpha(\lambda, L)$ is the absorptivity of the solar cell with an active layer thickness of $L$.

According to the detailed balance theory and Boltzmann statistics, the radiative current $J_e$ can be expressed as

$$J_e = J_0 \left[ \exp\left(\frac{qV}{k_B T}\right) - 1 \right] \quad (S7)$$

where $J_0$ is the radiative saturation current, which can be calculated by the blackbody radiation law.

$$J_0 = q \int_0^\infty \alpha(\lambda, L) \frac{\Gamma_0(\lambda)\lambda}{hc_0} d\lambda$$

$$\Gamma_0(\lambda) = \int_0^{2\pi} d\varphi \int_0^{\theta_m} S(\lambda) \cos(\theta) \sin(\theta) d\theta = \pi \sin^2\theta_m(\lambda) S(\lambda)$$

$$S(\lambda) = \frac{2hc_0^2}{\lambda^5} \frac{1}{\exp\left(\frac{hc_0}{\lambda k_B T}\right) - 1} \quad (S8)$$

where $\Gamma_0(\lambda)$ is the blackbody emission spectrum of the solar cell and $S(\lambda)$ is the thermal radiance of the cell. $\theta_m$ is the maximum emission angle related to the angular restriction design.

The detailed balance model is an indispensible tool for predicting the efficiency limit of PVSCs. The model only considers the detailed balance between absorbed photons and emitted photons. However, for the practical PVSCs, the inclusions of the non-radiative recombination and the influence of carrier transport layers are important for understanding device physics qualitatively and unveiling energy loss quantitatively.

In our revised detailed balance model, we include the currents $J_r$ and $J_n$ induced by the radiative and non-radiative recombination, respectively. Moreover, we introduce the series resistance $R_s$ that describes the ohmic loss from the contacts, carrier transport layers, and the hetero-junction interfaces between the perovskite and carrier transport layers. The introduced shunt resistance $R_{sh}$ captures the defects and voids induced current leakage. The revised detailed balance model captures different recombination mechanisms, the influences of carrier transport layers and contacts, photon recycling effects, and optical effects (light trapping). Compared to the circuit model, only three parameters are needed to be extracted. Thus, it is a promising tool to photovoltaic science and engineering.

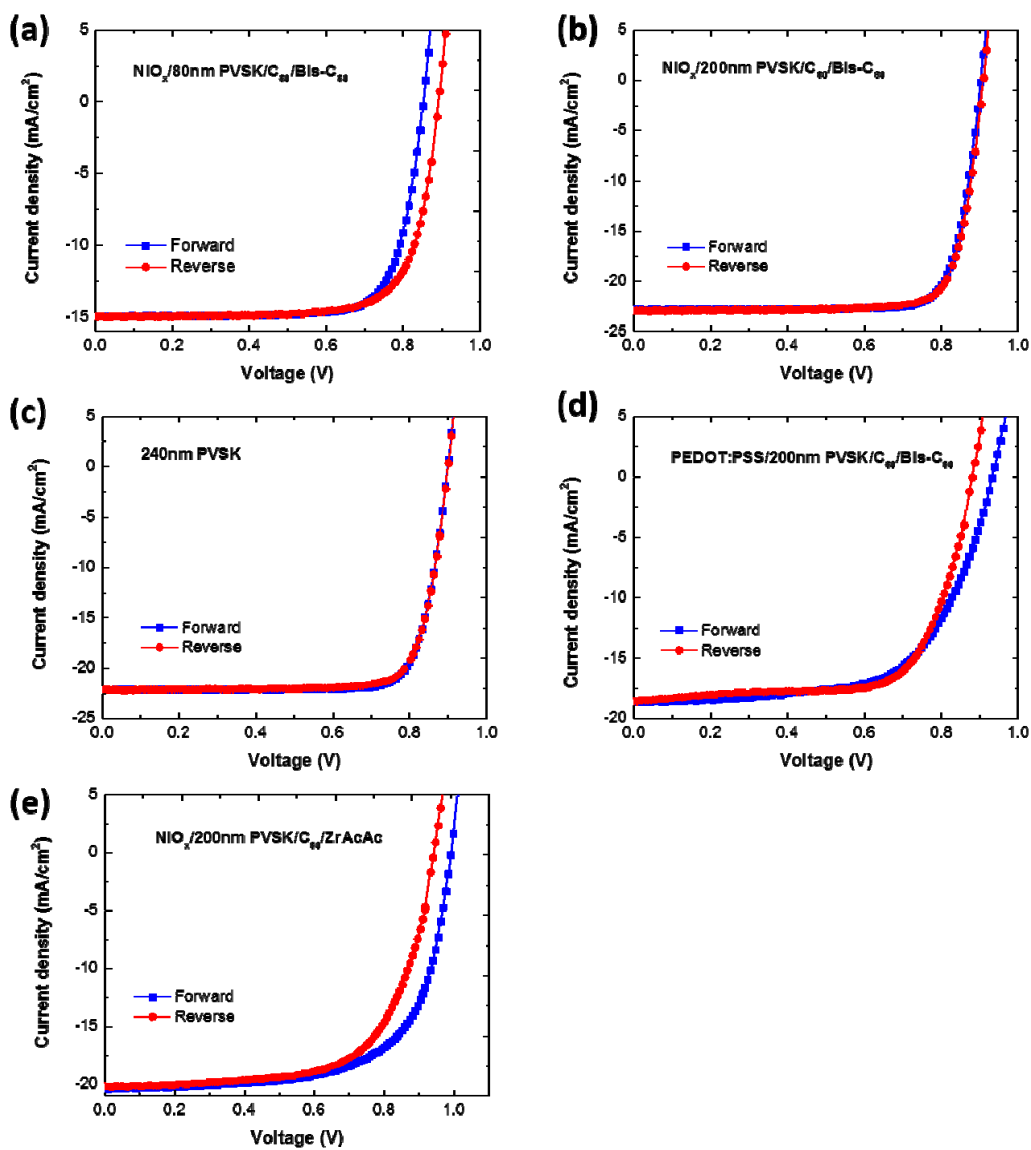

**Figure S1**. The *J-V* characteristics for perovskite solar cells with different perovskite thickness (a, b, and c) and carrier transport layers (b, d, and e) measured under different scan directions.

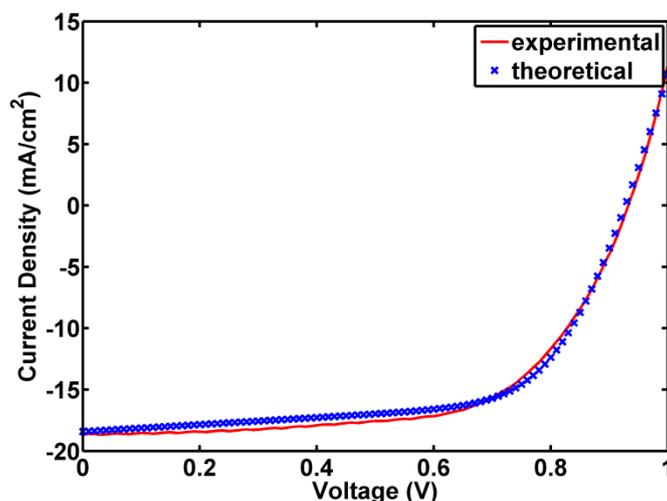

**Figure S2**. The theoretically fitted and experimentally measured *J-V* characteristics for perovskite solar cells incorporating PEDOT:PSS and Bis-$C_{60}$ as the hole transport layer and interfacial layer, respectively. The device structure is given as: ITO/ PEDOT:PSS (40 nm)/ $CH_3NH_3PbI_3$ (200 nm)/ $C_{60}$ (50 nm)/ Bis-$C_{60}$ (10 nm)/ Ag (120 nm). The forward scanning is employed. The relative fitting error is 3.34%.

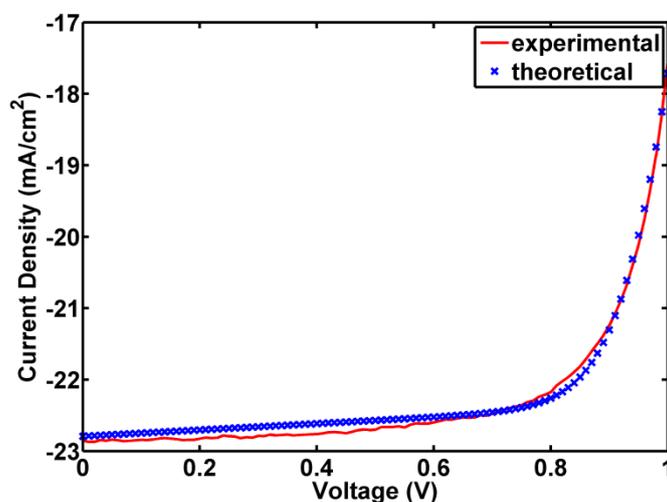

**Figure S3**. The theoretically fitted and experimentally measured *J-V* characteristics for the high-performance perovskite solar cell. The device structure is given as: ITO/ $NiO_x$ (20 nm)/ $CH_3NH_3PbI_3(Cl)$ (300 nm)/ PCBM:$C_{60}$ mixture (50 nm)/ Bis-$C_{60}$ (10 nm)/ Ag (120 nm). The relative fitting error is 0.48%.